\documentclass[12pt]{article}
\usepackage[margin=1in]{geometry}
\usepackage{amsthm,amssymb,mathrsfs,setspace,amsmath,latexsym,footmisc,graphicx,booktabs}
\usepackage{color}
\usepackage{textcomp}
\usepackage{multirow} 
\usepackage{makecell}
\usepackage{subfig}
\usepackage{enumitem}
\usepackage{hyperref}

\theoremstyle{plain}

\theoremstyle{definition}

\theoremstyle{remark}

\numberwithin{equation}{section}

\begin{document}
\title{\sc A Model for Censored Reliability Data with Two Dependent Failure Modes and Prediction of Future Failures}
\author{Aakash Agrawal \thanks{Independent Researcher, Ambikapur, Chhattisgarh 497001, India; Email: akash2016@alumni.iitg.ac.in} ,
Debanjan Mitra \thanks{Indian Institute of Management Udaipur, Rajasthan 313001,
India; Email: debanjan.mitra@iimu.ac.in; Telephone: +91-294-2477130 (Office)} , and Ayon Ganguly \thanks{Indian
Institute of Technology Guwahati, Assam 781039, India; Email:
aganguly@iitg.ac.in}}
\date{}
\maketitle

\begin{abstract}
     Quite often, we observe reliability data with two failure modes that may influence each other, resulting in a setting of dependent failure modes. Here, we discuss modelling of censored reliability data with two dependent failure modes by using a bivariate Weibull model with distinct shape parameters which we construct as an extension of the well-known Marshall-Olkin bivariate exponential model in reliability. Likelihood inference for modelling censored reliability data with two dependent failure modes by using the proposed bivariate Weibull distribution with distinct shape parameters is discussed. Bayesian analysis for this issue is also discussed. Through a Monte Carlo simulation study, the proposed methods of inference are observed to provide satisfactory results. A problem of practical interest for reliability engineers is to predict field failures of units at a future time. Frequentist and Bayesian methods for prediction of future failures are developed in this setting of censored reliability data with two dependent failure modes. An illustrative example based on a real data on device failure with two failure modes is presented. The model and methodology presented in this article provide a complete and comprehensive treatment of modelling censored reliability data with two dependent failure modes, and address some practical prediction issues. 
\end{abstract}



\doublespacing
\section{\sc Introduction}
In reliability studies, quite often, more than one failure mode is observed to be present. In reliability and survival analysis, this scenario is known as competing risks, the failure modes being termed as the risk factors that compete among each other; see Meeker and Escobar~\cite{Meeker}, and Crowder~\cite{Crowder} for more details. For example, consider the data presented in 
Meeker and Escobar~\cite{Meeker}, pertaining to lifetimes of a device which is part of a larger system. For a sample of this device, there are two failure modes: Mode S and Mode W. Mode S failures are caused by accumulated damage from power-line voltage spikes, while Mode W failures are caused by normal product wear. Naturally for a unit, a Mode S failure would preclude a Mode W failure and vice versa.    

The early literature in reliability and survival analysis on competing risks was developed under the latent failure time assumption~\cite{Cox}. In this approach, lifetimes corresponding to different failure modes or risk factors are modelled marginally, independently of each other. There are several lifetime distributions available in literature that are particularly useful for reliability data because of their specific characteristics; these distributions may be used as marginal failure-time models corresponding to different independent failure modes. Apart from the classical models, some such failure-time models that have been recently discussed are additive Weibull of Xie and Lai~\cite{Xie}, modified Weibull of Almalki and Yuan~\cite{Almalki}, and additive modified Weibull of He et al.~\cite{He}; see also Abba et al.~\cite{Abba} and the references therein.

However in reality, more often than not, the failure modes influence each other by some complex mechanisms; for the real data example on device failure mentioned above, Mode S and Mode W failures may be related by some inherent yet unknown mechanism. Thus, generally speaking, considering marginal independent failure-time models for different failure modes would be a rather simplistic approach in such cases. Instead, a more reasonable approach in such cases would be to consider a joint model for the failure-times corresponding to different failure modes, leveraging the dependence structure of the joint model to capture the dependence among the failure modes. Bivariate models are particularly of interest when there are two failure modes, such as Mode S and Mode W failures in the device failure data mentioned above. Lawless~\cite{Lawless} constructed the general likelihood function for modelling of this type.



Recently, some bivariate models have been used in reliability literature to capture dependence between the lifetimes of two components; see, for example, Fan and Hsu~\cite{Fan}, and Oliviera et al.~\cite{Oliviera}. In Oliviera et al.~\cite{Oliviera}, a Marshall-Olkin type bivariate distribution was consiered to acheieve some generalised results. The Marshall-Olkin bivariate exponential (MOBE) model~\cite{Marshall67} is considered as the reference model in numerous studies in reliability. The MOBE model originates from a shock model that assumes independent occurrences of shocks to the components of a two-component series system. One of the most interesting features of this model is that it has a component of singularity that makes it particularly suitable for modelling bivariate reliability data with ties. 

A practical difficulty in using the MOBE distribution in reliability modelling however is that the marginals in this case are exponential distributions. As the failure rate (also known as hazard rate) of an exponential distribution is constant, the MOBE model is not suitable for data with marginals exhibiting non-constant failure rates. In reality, bathtub-shape failure rate functions characterize lifetimes of many reliability systems, and a constant failure rate function as observed in exponential distribution is not ideal for modelling purposes. Marshall and Olkin~\cite{Marshall67} gave a theoretical outline of an extension of the bivariate exponential model to a bivariate Weibull model, although they did not explore this bivariate Weibull model in detail. A bivariate Weibull model having univariate Weibull models as the marginals, would be suitable for modelling reliability data with non-constant marginal failure rates. 

In this paper, we develop a comprehensive inferential framework for modelling censored reliability data with two dependent failure modes by using a bivariate Weibull distribution which we construct by considering an extension of the MOBE model by means of distinct shape parameters. The bivariate Weibull model considered here, has two distinct shape parameters added to the well-known MOBE distribtion. The resulting model, a bivariate Weibull distribution with distinct shape parameters and Weibull marginals, is much more flexible in handling different types of reliability data showing various types of marginal patters for failure rates. It may be noted here that an extension of the MOBE distribution with common shape parameters and its applications in competing risks have been considered in literature; see for example, Kundu and Dey~\cite{Kundu-EM}, Kundu and Gupta~\cite{Kundu-Bayes}, Feizvidian and Hashemi~\cite{Feiz2015}, and Samanta and Kundu~\cite{Samanta}. The assumption of a common shape parameter for the component lifetimes of a series system greatly reduces complexity. However, for practical applications, the assumption of a common shape parameter for distributions of the component lifetimes of a two-component series system is quite restrictive, and may not have any physical justification.  

The bivariate Weibull model we consider in this paper is constructed without any such restrictive assumption on the shape parameters of the distributions of component lifetimes of a two-component series system. Physically, this amounts to assuming Weibull distributions with distinct shape parameters for the independent shocks that give rise to this model, generalizing the shock model based on exponential distributions as assumed by Marshall and Olkin~\cite{Marshall67}. We call this model the Marshall-Olkin bivariate Weibull distribution with distinct shape parameters; for convenience and brevity, this model is referred to as the MOBWDS model in this paper. The proposed model is then used for modelling censored reliability data with two dependent failure modes. 

The main contributions of this paper are as follows. First, the MOBWDS distribution as a general extension of the MOBE distribution is constructed; to the best of our knowledge, such an extension has not been considered in reliability literature so far. Then, the proposed MOBWDS distribution is used to model censored reliability data with two dependent failure modes. Likelihood and Bayesian inferential methods for modelling censored reliability data with dependent failure modes are discussed in detail, and the performances of the inferential methods are assessed through a detailed Monte Carlo simulation study. A case study based on a real data on device failure is presented as an illustration. 

Reliability engineers often need to predict future failures of units for predictive maintenance purposes. Generally speaking, within-sample prediction of future failures is of practical relevance; in within-sample prediction, based on past data from the failure process, future failures of units already in test are predicted~\cite{Lewis-Beck}. In reliability literature, many researchers have addressed various prediction issues. For example, Lewis-Beck et al.~\cite{Lewis-Beck} developed approaches for predicting future failures based on heterogeneous reliability field data, see also Escobar and Meeker~\cite{Escobar} for a general discussion on statistical aspects of prediction based on censored data. Recently, failure prediction approaches based on model-based teachnique and machine learning tools have also been discussed (\cite{Rifaai}, \cite{Fan-X}).
 
Prediction of future failures is complicated when two dependent failure modes are present. We develop frequentist as well as Bayesian methods for predicting future failures in this setting of censored reliability data with two dependent faiilure modes, and provide illustrative example of the same based on the device failure data. This is another main contribution of this paper. In summary, this paper provides a comprehensive approach for modelling censored reliability data with dependent failure modes, and address some relevant prediction issues.  

The paper is organized as follows. 
Section \ref{sec:MOBWDS} presents construction of the MOBWDS model. In Section \ref{sec:modelling}, likelihood inference for modelling reliability data with dependent failure modes by the proposed MOBWDS model is discussed. Bayesian inference using Markov Chain Monte Carlo technique for this issue is developed in Section \ref{sec:Bayes}. Results of a detailed simulation study are presented in Section \ref{sec:MC}. Some relevant and practical prediction issues are discussed in Section \ref{sec:Pred}. An illustrative example of all the proposed methods of inference and prediction based on a real data on device failure is provided in Section \ref{sec:Data-analysis}. Finally, the paper is concluded in Section \ref{sec:Conclusion} with some remarks. 

\section{\sc The bivariate Weibull model with distinct shape parameters}\label{sec:MOBWDS}
The cumulative distribution function (CDF) of the univariate Weibull distribution We$(\alpha, \lambda)$ is given by 
\begin{equation}
F(t; \alpha, \lambda) = 1 - e^{-\lambda t^{\alpha}}, \quad t>0, \nonumber
\end{equation}
where $\alpha (>0)$ and $\lambda (>0)$ are the shape and scale parameters, respectively. The MOBWDS distribution with distinct shape parameters is constructed as follows. Consider three independent random variables $U_0$, $U_1$, $U_2$, with 
$$
U_i \sim \textrm{We}(\alpha_i, \lambda) \quad i=1,2,3.
$$
Define, 
$$X = min(U_1, U_0), \quad \textrm{and} \quad Y = min(U_2, U_0).$$ Then, the random vector $(X, Y)$ follows the MOBWDS distribution with distinct shape parameters, and we denote it by
$$(X, Y) \sim \textrm{MOBWDS}(\alpha_0, \alpha_1, \alpha_2, \lambda),$$ where $\alpha_0, \alpha_1, \alpha_2$ are the shape parameters, and $\lambda$ is a scale parameter. The survival function of the distribution is given by 
\begin{align*}
S_{\textrm{MOBWDS}}(x, y) & = P(X \geq x, Y \geq y) \\
			   & = P(U_1 \geq x)  P(U_2 \geq y) P(U_0 \geq z),
\end{align*}
where $z = max(x, y)$. Using the distributions of $U_0$, $U_1$, and $U_2$, we have 
\begin{align}
   {S_{\textrm{MOBWDS}}(x, y)} =\begin{cases}
      S_1(x, y), & \text{for} \quad 0 < x < y < \infty \\
      S_2(x, y), & \text{for} \quad 0 < y < x < \infty \\
      S_0(x), & \text{for} \quad 0 < x = y < \infty, 
   \end{cases} \label{eq:surv}
\end{align}
where 
$$
S_1(x, y) = e^{-\lambda(x^{\alpha_1}+y^{\alpha_2}+y^{\alpha_0})},
$$
$$
S_2(x, y) = e^{-\lambda(y^{\alpha_2}+x^{\alpha_1}+x^{\alpha_0})},
$$
$$
S_0(x) = e^{-\lambda(x^{\alpha_1}+x^{\alpha_2}+x^{\alpha_0})}.
$$

For cases other than $0 < x = y < \infty$, the joint density of $X$ and $Y$
can be derived from $S_{\textrm{MOBWDS}}(x, y)$, by finding $\frac{\partial^2}{\partial x \partial y}S_{\textrm{MOBWDS}}(x,y)$. When $0 < x = y < \infty$, closed form expression of the joint density is not available. That is, the density of $(X,Y)$ for the MOBWDS distribution is given by
\begin{align}
   {f_{\textrm{MOBWDS}}(x, y)} =\begin{cases}
      f_1(x, y), & \text{for} \quad 0 < x < y < \infty \\
      f_2(x,y),  & \text{for} \quad 0 < y < x < \infty \\
      f_0(x),    & \text{for} \quad 0 < x = y < \infty,
   \end{cases} \label{eq:pdf}
\end{align}
where 
$$
f_1(x,y) = \frac{\partial^2}{\partial x \partial y}S_1(x,y) = \lambda^2\alpha_1x^{\alpha_1-1}(\alpha_2 y^{\alpha_2-1} + \alpha_0 y^{\alpha_0-1}) e^{-\lambda(x^{\alpha_1} + y^{\alpha_2} + y^{\alpha_0})},
$$
$$
f_2(x,y) = \frac{\partial^2}{\partial x \partial y}S_2(x,y) = \lambda^2\alpha_2y^{\alpha_2-1}(\alpha_1 x^{\alpha_1-1} + \alpha_0 x^{\alpha_0-1}) e^{-\lambda(y^{\alpha_2} + x^{\alpha_1} + x^{\alpha_0})},
$$
and
$$
f_0(x) = P(X=Y)\bigg[-\frac{\partial}{\partial x}S_0(x)\bigg] = P(X=Y)\lambda(\alpha_1x^{\alpha_1-1}+\alpha_2x^{\alpha_2-1}+\alpha_0x^{\alpha_0-1})e^{-\lambda(x^{\alpha_1}+x^{\alpha_2}+x^{\alpha_0})},
$$
are obtained using Eq.\eqref{eq:surv}. 

Also note that 
\begin{equation}
\int_0^{\infty}\int_x^{\infty} f_1(x,y)dydx + \int_0^{\infty}\int_y^{\infty} f_2(x,y)dxdy + \int_0^{\infty}f_0(x) = 1. \label{eq:sum-dens}
\end{equation}
Here, 
\begin{equation}
\int_0^{\infty}\int_x^{\infty} f_1(x,y)dydx = \int_0^{\infty} \lambda \alpha_1 x^{\alpha_1 - 1} e^{-\lambda(x^{\alpha_0}+x^{\alpha_1}+x^{\alpha_2})} dx, \label{eq:f-1}
\end{equation}
and 
\begin{equation}
\int_0^{\infty}\int_y^{\infty} f_2(x,y)dxdy = \int_0^{\infty} \lambda \alpha_2 y^{\alpha_2 -1} e^{-\lambda(y^{\alpha_0}+y^{\alpha_1}+y^{\alpha_2})} dy. \label{eq:f-2}
\end{equation}
From Eqs.\eqref{eq:sum-dens}, \eqref{eq:f-1}, and \eqref{eq:f-2}, it can be readily seen that $f_0(x)$ does not have a closed form expression. Here, $P(X = Y)$ can be obtained to be 
\begin{equation}
P(X = Y) = 1 - \lambda \int_0^{\infty}(\alpha_1 x^{\alpha_1-1} + \alpha_2x^{\alpha_2-1})e^{-\lambda(x^{\alpha_0}+x^{\alpha_1}+x^{\alpha_2})}dx. \label{eq:sing-prob}
\end{equation}

\begin{figure}
	\centering
	\subfloat{\includegraphics[width=.4\linewidth, height=.4\linewidth]{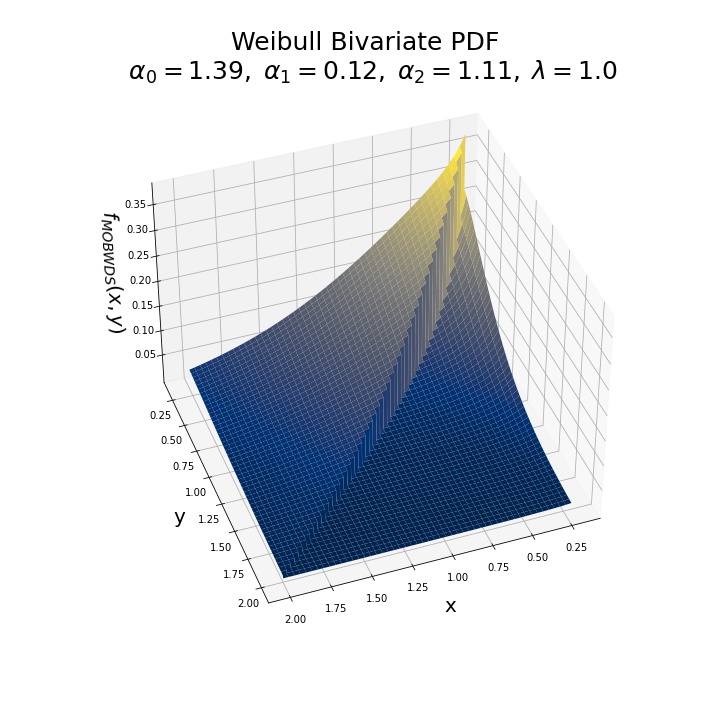}}
	\subfloat{\includegraphics[width=.4\linewidth, height=.4\linewidth]{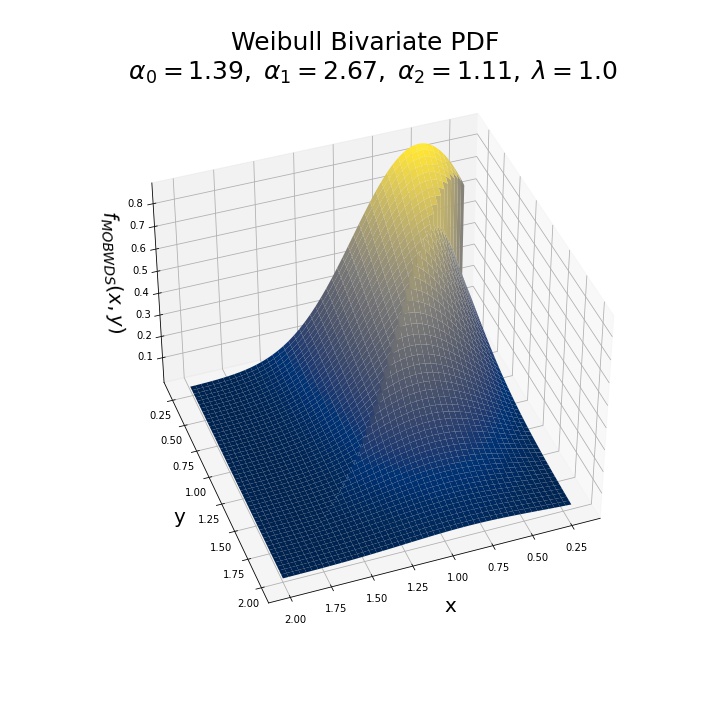}}
	\hfill
	\subfloat{\includegraphics[width=.4\linewidth, height=.4\linewidth]{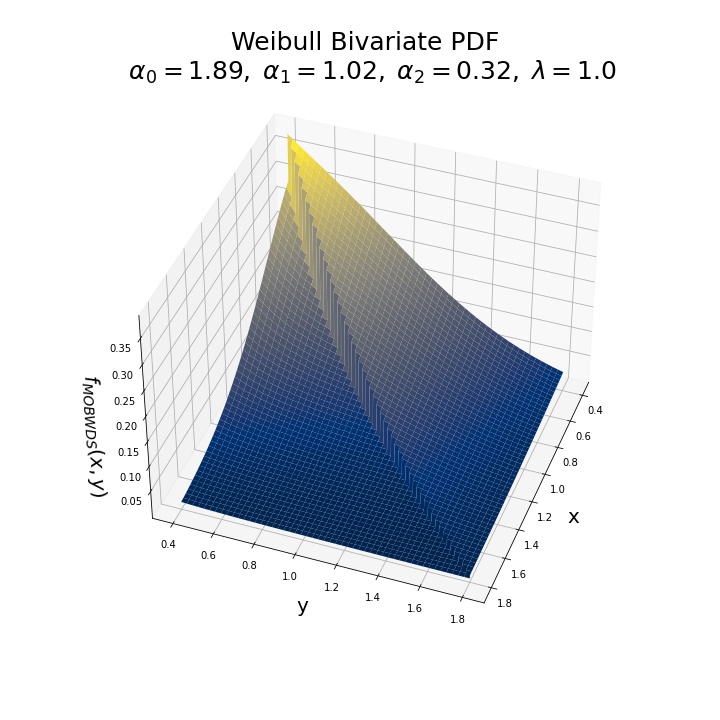}}
	\subfloat{\includegraphics[width=.4\linewidth, height=.4\linewidth]{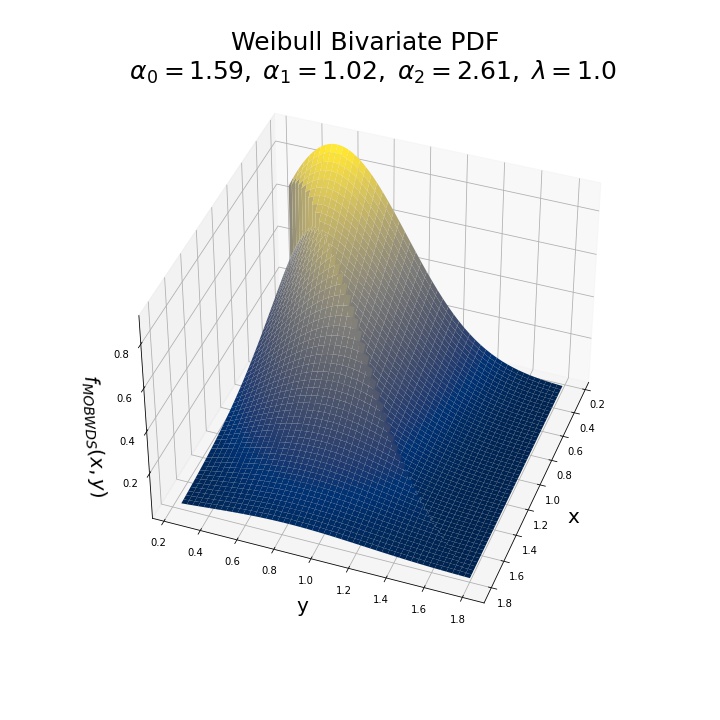}}
	\hfill
	\subfloat{\includegraphics[width=.4\linewidth, height=.4\linewidth]{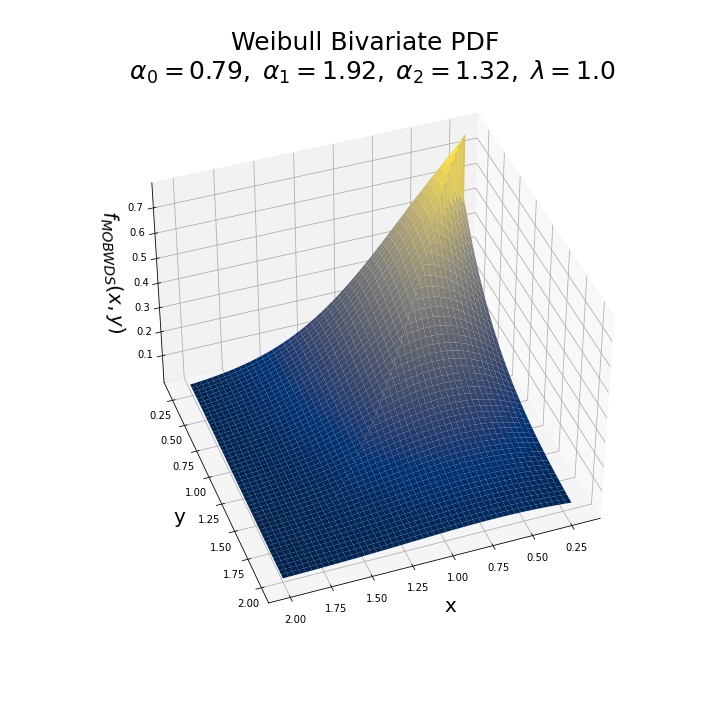}}
	\subfloat{\includegraphics[width=.4\linewidth, height=.4\linewidth]{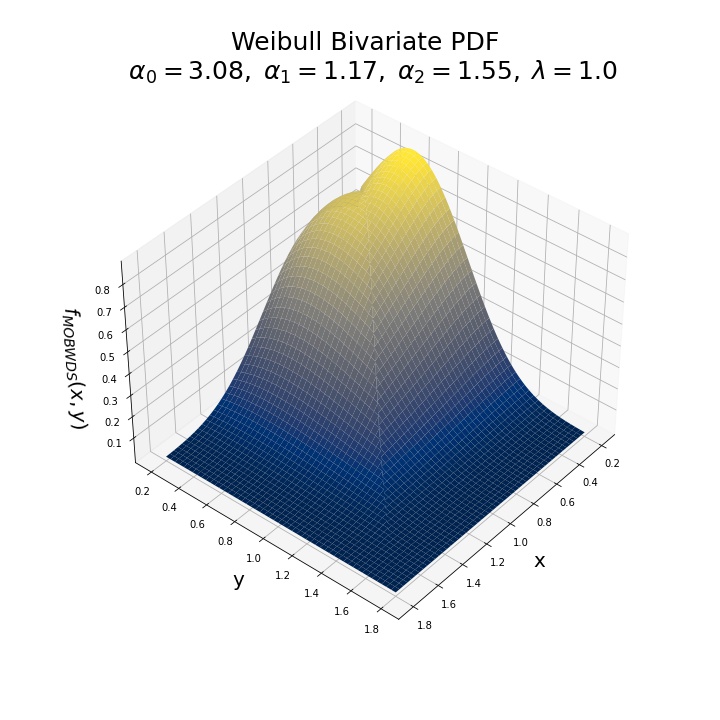}}
	\caption{PDF of the MOBWDS$(\alpha_0, \alpha_1, \alpha_2, \lambda)$ distribution for different values of parameters.}
	\label{dens-3D}
\end{figure}

In Figure \ref{dens-3D}, the PDF of the MOBWDS distribution with distinct shape parameters is presented for different values of the parameters. 



\section{\sc Likelihood inference for reliability data with two dependent failure modes}\label{sec:modelling}
Consider a reliability study (field or laboratory) with $n$ units. Suppose there are two failure modes, called Mode 1 and Mode 2, for failures of study units. The failure modes may be dependent on each other. The lifetimes of each unit in the study can then be considered as a bivariate random vector. Our approach for modelling the data would involve the use of a bivariate model for the joint lifetimes corresponding to the two failure modes. That is, if $X$ and $Y$ denote the lifetimes corresponding to Mode 1 and 2, respectively, then we would consider a bivariate model for $X$ and $Y$. Suppose the joint survival function (SF) of $X$ and $Y$ is denoted by $S_{X,Y}(\cdot, \cdot)$, and the corresponding joint probability density function (PDF) is denoted by $f_{X,Y}(\cdot, \cdot)$.    

For each observed failure, there are three possibilities: (a) it is a Mode 1 failure, (b) it is a Mode 2 failure, and (c) the failure is a tie between Mode 1 and Mode 2 (i.e., failures from both modes at the same time). Define the following indicator variables: 
\begin{eqnarray}
\delta_1 =
\begin{cases}
1, & \textrm{if it is a Mode 1 failure} \\ 
0, & \textrm{otherwise}, \nonumber
\end{cases}
\end{eqnarray}
\begin{eqnarray}
\delta_2 =
\begin{cases}
1, & \textrm{if it is a Mode 2 failure} \\ 
0, & \textrm{otherwise}, \nonumber
\end{cases}
\end{eqnarray}
and 
\begin{eqnarray}
\delta_0 =
\begin{cases}
1, & \textrm{if it is a tie between Mode 1 and 2} \\ 
0, & \textrm{otherwise}. \nonumber
\end{cases}
\end{eqnarray}
Note that when $\delta_0 = \delta_1 = \delta_2$ = 1 for a unit implies that the unit has not failed during the study. Such units are called the right censored units in reliability and survival analyses. Thus, the observed data in this case is of the format
$$
Data = (t_i, \delta_{i0}, \delta_{i1}, \delta_{i2}), \quad i=1,...,n, 
$$
where $t_i$ is the lifetime of the $i-$th unit; $t_i$ could be an observed failure, or a right censored lifetime.  

A failed unit with lifetime $t$ and $\delta_1=1$ (Mode 1 failure) will contribute $\left.-\frac{\partial}{\partial x} S_{1}(x, y)\right|_{x=t, y=t}$ to the likelihood. Similarly, the contribution to the likelihood by a failed unit with lifetime $t$ and $\delta_2=1$ (Mode 2 failure) will be $-\left.\frac{\partial}{\partial y} S_{2}(x, y)\right|_{x=t, y=t}$. For each failure that is a result of a tie between the two modes (i.e., $\delta_0 = 1$), the contribution to the likelihood will be $f_0(t)$. For each of right censored units, the contribution to the likelihood will be $S_0\left(t\right)$. Combining these cases, the observed likelihood is given by
\begin{eqnarray}
& L(\boldsymbol \theta) & \propto \prod_{i=1}^{n}\left[f_{0}\left(t_{i}\right)\right]^{\delta_{i 0}}\times \left[-\left.\frac{\partial}{\partial x} S_{1}(x, y)\right|_{x=t_{i}, y=t_{i}}\right]^{\delta_{i 1}} \nonumber \\
&& \times \left[-\left.\frac{\partial}{\partial y} S_{2}(x, y)\right|_{x=t_{i}, y=t_{i}}\right]^{\delta_{i 2}} \times  \left[S_{0}\left(t_{i}\right)\right]^{1-\delta_{i 0}-\delta_{i 1}-\delta_{i 2}},
\label{eq:LL}
\end{eqnarray}
where $\boldsymbol \theta$ is the vector of relevant model parameters.

Suppose that  
$$
(X, Y) \sim \textrm{MOBWDS}(\alpha_0, \alpha_1, \alpha_2, \lambda).
$$
The SF and PDF of the MOBWDS distribution with distinct shape parameters are given in Eqs\eqref{eq:surv} and \eqref{eq:pdf}, respectively. 

Substituting the PDF and SF of the MOBWDS distribution in Eq.\eqref{eq:LL}, we have the likelihood under the MOBWDS model for censored reliability data with two dependent failure modes  as 
\begin{eqnarray}
& L(\boldsymbol \theta) & \propto  \alpha_1^{m_1} \alpha_2^{m_2} \lambda^{m_0+m_1+m_2} (P^*)^{m_0} e^{- \lambda \sum_{i=1}^{n}(t_i^{\alpha_0}+t_i^{\alpha_1}+t_i^{\alpha_2})} \nonumber \\
&& \prod_{i \in I_0}(\alpha_0t_i^{\alpha_0-1}+\alpha_1t_i^{\alpha_1-1}+\alpha_2t_i^{\alpha_2-1}) \prod_{i \in I_1} t_i^{\alpha_1 - 1} \prod_{i \in I_2} t_i^{\alpha_2 - 1}, \label{eq:Likelihood}
\end{eqnarray}
where $\boldsymbol \theta = (\alpha_0, \alpha_1, \alpha_2, \lambda)$, and $P^* = P(X = Y)$ as given in Eq.\eqref{eq:sing-prob}.

Define index sets $I_j$, $j=0,1,2$ as follows: 
\[
I_j = \big\{i: \delta_{ij}=1, i=1,...,n\big\}, \quad j=0,1,2.
\]
Also, denote $|I_1| = m_1$ (i.e., number of failures from Mode 1), $|I_2| = m_2$ (i.e., number of failures from Mode 2), and $|I_0| = m_0$ (i.e., number of failures from both modes). Then, a simplified form of the log-likelihood function in this case is given by
\begin{eqnarray}
& \log L(\boldsymbol \theta) & = m_0 \log P^* + m_1 \log \alpha_1 + m_2 \log \alpha_2 + (m_0+m_1+m_2)\log \lambda \nonumber \\
&& - \lambda \sum_{i=1}^{n}(t_i^{\alpha_0}+t_i^{\alpha_1}+t_i^{\alpha_2}) + \sum_{i \in I_0}\log(\alpha_0t_i^{\alpha_0-1}+\alpha_1t_i^{\alpha_1-1}+\alpha_2t_i^{\alpha_2-1}) \nonumber \\
&& + (\alpha_1-1) \sum_{i \in I_1} \log(t_i) + (\alpha_2-1) \sum_{i \in I_2} \log(t_i). \label{Loglik}
\end{eqnarray}
The maximum likelihood estimates (MLEs) of the model parameters can be obtained by maximizing Eq.\eqref{Loglik}, by the help of a numerical technique. Standard statistical software offer routine optimizers; for example, the \texttt{R} software has a routine optimizer \texttt{optim()} that performs general purpose optimization by using Nelder--Mead method. Such routine optimizers can be used for numerically optimizing the log-likelihood function in Eq.\eqref{Loglik} to obtain the MLEs of model parameters.  

Along with the MLEs, it is also possible to provide 95\% confidence intervals for the parameters which can be obtained by using the observed Fisher information matrix. The observed Fisher information matrix is constructed by using the negative of the second derivatives of log-likelihood function in Eq.\eqref{Loglik} with respect to the parameters. Let $\boldsymbol J(\boldsymbol \theta)$ denote the observed Fisher information matrix for $\boldsymbol \theta$. We have $\boldsymbol J(\boldsymbol \theta)$ given by
$$\boldsymbol J(\boldsymbol \theta) = -\nabla^2(\log L(\boldsymbol \theta)),$$
where the $\nabla^2$ operator signifies second derivatives. Using asymptotic normality of the MLEs, we have 
$$
\sqrt{n}(\widehat{\boldsymbol \theta} - \boldsymbol \theta) \rightarrow \boldsymbol N_4(\boldsymbol 0, \boldsymbol J^{-1}(\boldsymbol \theta)|_{\widehat{\boldsymbol \theta} = \boldsymbol \theta}),
$$
$N_4(\boldsymbol \mu, \boldsymbol \Sigma)$ being a normal distribution of four dimensions with mean vector $\boldsymbol \mu$ and covariance matrix $\boldsymbol \Sigma$. Asymptotic 95\% confidence intervals (CIs) for the model parameters can be constructed easily by using this information; for example, for $\alpha_0$, the CI is given by  
$$\alpha_0 \pm 1.96\sqrt{(\boldsymbol J^{-1}(\boldsymbol \theta)|_{\widehat{\boldsymbol \theta} = \boldsymbol \theta})_{1,1}},
$$
where $(\boldsymbol J^{-1}(\boldsymbol \theta)|_{\widehat{\boldsymbol
\theta} = \boldsymbol \theta})_{i,j}$ is the $(i,j)-$th element of
$\boldsymbol J^{-1}(\boldsymbol \theta)|_{\widehat{\boldsymbol \theta} =
\boldsymbol \theta}$.  

\section{\sc Bayesian Analysis}\label{sec:Bayes}
\subsection{\sc Prior Assumptions}
As the likelihood inference does not result in closed form estimates of the model parameters, it is not possible to carry out exact inference based on the MLEs; one has to reply on asymptotic inference as described in the previous Section. Therefore, Bayesian inference with properly chosen prior distributions seems to be a natural alternative. 

The assumed prior distribution on $(\alpha_0, \alpha_1, \alpha_2)$ should be reflective of the dependence that we aim to capture in the two-failure mode model. For this reason, following Pena and Gupta~\cite{Pena}, a multivariate Gamma-Dirichlet prior with hyperparameters $a > 0$, $b > 0$, $a_0 > 0$, $a_1 > 0$, and $a_2 > 0$, denoted by $\textrm{GD}(a, b, a_0, a_1, a_2)$, is assumed on the shape parameters ($\alpha_0$, $\alpha_1$, $\alpha_2$). That is, the prior density $\pi_0(\boldsymbol \theta)$ is given by:
\begin{equation}
\pi_0(\boldsymbol \theta) = \pi_{0}\left(\alpha_{0}, \alpha_{1}, \alpha_{2} \mid a, b, a_{0}, a_{1}, a_{2}\right)=\frac{\Gamma(\bar{a})}{\Gamma(a)}(b \alpha)^{a-\bar{a}} \prod_{i=0}^{2} \frac{b^{a_{i}}}{\Gamma\left(a_{i}\right)} \alpha_{i}^{a_{i}-1} e^{-b \alpha_{i}}; \alpha_0, \alpha_1, \alpha_2 > 0, \nonumber
\end{equation}
where $\bar{a} = a_0 + a_1 + a_2$ and $\alpha = \alpha_0 + \alpha_1 + \alpha_2$. Note that the dependence among $\alpha_0$, $\alpha_1$ and $\alpha_2$ can be controlled through the hyper parameters of the Gamma-Dirichlet distribution $\textrm{GD}(a, b, a_0, a_1, a_2)$. When $a = \bar{a}$, the above model implies that the shape parameters $\alpha_0$, $\alpha_1$ and $\alpha_2$ have independent gamma priors. 

The prior distribution on parameter $\lambda$ is assumed to be gamma, with hyperparameters $c_1 > 0$ and $c_2 > 0$, denoted by $\textrm{GA}(c_1, c_2)$. The priors on $\lambda$ and $\alpha_0, \alpha_1$ and $\alpha_2$ are assumed independent. Therefore, the joint prior on ($\alpha_0$, $\alpha_1$, $\alpha_2$, $\lambda$) is given by:
\begin{equation}
\pi_{1}\left(\alpha_{0}, \alpha_{1}, \alpha_{2}, \lambda \mid a, b, a_{0}, a_{1}, a_{2}, c_{1}, c_{2}\right)=\frac{c_{1}^{c_{2}}}{\Gamma\left(c_{2}\right)} e^{-c_{1} \lambda} \lambda^{c_{2}-1} \times \frac{\Gamma(\bar{a})}{\Gamma(a)}(b \alpha)^{a-\bar{a}} \prod_{i=0}^{2} \frac{b^{a_{i}}}{\Gamma\left(a_{i}\right)} \alpha_{i}^{a_{i}-1} e^{-b \alpha_{i}}. \label{eq:prior}
\end{equation}

The joint prior given in Eq.\eqref{eq:prior} is a very flexible one as it can capture and control the dependence among the parameters $\alpha_0, \alpha_1$ and $\alpha_2$. It may be noted here that in literature, the Gamma-Dirichlet prior has been assumed on scale parameters~\cite{Samanta}. In this research, we have assumed the Gamma-Dirichlet prior on the shape parameters and this approach has not been explored by any researcher to the best of our knowledge. 

\subsection{\sc Posterior analysis: Metropolis Hastings algorithm}
Using the likelihood function in Eq.\eqref{eq:Likelihood} and the prior assumption in Eq.\eqref{eq:prior}, the posterior distribution of ($\alpha_0$, $\alpha_1$, $\alpha_2$, $\lambda$) is obtained as
\begin{equation}
\tilde{\pi}\left(\boldsymbol \theta \mid \text {Data}\right) = \tilde{\pi}\left(\alpha_0, \alpha_1, \alpha_2, \lambda \mid \text {Data}\right) \propto L(\boldsymbol \theta) \pi_{1}(\alpha_0, \alpha_1, \alpha_2, \lambda). \label{eq:posterior}
\end{equation}
The Bayes estimate of a parametric function $g(\boldsymbol \theta)$ with respect to squared error loss is thus given by
\begin{eqnarray}
& \widehat{g}_{Bayes}(\boldsymbol \theta) & = \int_{\boldsymbol \theta}g(\boldsymbol \theta) \tilde{\pi}\left(\boldsymbol \theta \mid \text {Data}\right) d\boldsymbol \theta \nonumber \\
&&= \int_0^{\infty}\int_{0}^{\infty}\int_0^{\infty}\int_0^{\infty}g(\boldsymbol \theta) \tilde{\pi}\left(\alpha_0, \alpha_1, \alpha_2, \lambda \mid \text {Data}\right) d\alpha_0 d\alpha_1 d\alpha_2 d\lambda.
\end{eqnarray}
Clearly, the Bayes estimate is not available in closed form, and we need to implement an approach based on Markov Chain Monte Carlo (MCMC) techniques. 
 
Here, we use the Metropolis-Hastings algorithm to obtain Bayes estimates and credible intervals for the parameters $\alpha_0$, $\alpha_1$, $\alpha_2$, and $\lambda$. Since all the parameters have a non-negative support, a multivariate folded normal distribution is used as the proposal distribution in the Metropolis-Hastings algorithm. In particular, the density function of a multivariate folded normal distribution with parameters $\boldsymbol \mu = (\mu_1, \mu_2, \mu_3, \mu_4) \in \mathbb{R}^4$ and positive definite matrix $\boldsymbol \Sigma$ given by 
$$
\boldsymbol \Sigma = \begin{bmatrix}
  \sigma_1^2 & 0 & 0 & 0\\ 
  0 & \sigma_2^2 & 0 & 0 \\
  0 & 0 & \sigma_3^2 & 0 \\
  0 & 0 & 0 & \sigma_4^2 
\end{bmatrix},
$$
evaluated at $\boldsymbol x = (x_1, x_2, x_3, x_4) \in \mathbb{R}^4$ is given by 
$$
Q(\boldsymbol x; \boldsymbol \mu, \boldsymbol \Sigma) = \frac{1}{\sigma_i \sqrt{2 \pi}} e^{-\frac{(x_i-\mu_i)^{2}}{2 \sigma_i^{2}}}+\frac{1}{\sigma_i \sqrt{2 \pi}} e^{-\frac{(x_i+\mu_i)^{2}}{2 \sigma_i^{2}}}.
$$ 

The Metropolis-Hastings algorithm at the $l$-th stage is as follows: \newline
\noindent {\sc Algorithm 1:} 
\begin{enumerate}
[label=\arabic*.,itemsep=-1ex]
   \item Step 1: The currently available value of the parameter vector is $\boldsymbol \theta^{(l)} = (\alpha_0^{(l)}, \alpha_1^{(l)}, \alpha_2^{(l)}, \lambda^{(l)})$.
   \item Step 2: Given $\boldsymbol \theta^{(l)}$, choose a proposed value $\boldsymbol \theta_p$ from the folded normal distribution with $\boldsymbol \mu = \boldsymbol \theta^{(l)}$ and $\boldsymbol \Sigma = \frac{1}{2}\boldsymbol I_4$ where $\boldsymbol I_4$ is the $4 \times 4$ identity matrix. 
   \item Step 3: Compute the acceptance probability $R$: 
   $$
R=\min \left(1, \frac{\tilde{\pi}(\boldsymbol \theta_p) Q\left(\boldsymbol \theta^{(l)} ; \boldsymbol \theta_p, \boldsymbol \Sigma\right)}{\tilde{\pi}\left(\boldsymbol \theta^{(l)}\right) Q\left(\boldsymbol \theta_p ; \boldsymbol \theta^{(l)}, \boldsymbol \Sigma\right)}\right)
$$
	As for the folded normal distribution, $Q(\boldsymbol \theta ; \boldsymbol \mu, \boldsymbol \Sigma) = Q(\boldsymbol \mu ; \boldsymbol \theta, \boldsymbol \Sigma)$, we can re-write acceptance probability as
	$$
R=\min \left(1, \frac{\tilde{\pi}(\boldsymbol \theta_p)}{\tilde{\pi}\left(\boldsymbol \theta^{(l)}\right)}\right)
$$
	\item Step 4: Take $\boldsymbol \theta^{(l+1)} = \boldsymbol \theta_p$ with probability $R$, and $\boldsymbol \theta^{(l+1)} = \boldsymbol \theta^{(l)}$ with probability $1-R$
	\item Step 5: Repeat steps 2-4 for the desired chain length, say $N$; $N = n_{burn} + n$ where $n_{burn}$ is the burn-in sample, and $n$ is the effective sample, and store $\boldsymbol \theta$ values after discarding the burn-in as samples from the posterior distribution:\\
	$\boldsymbol \theta^{(1)} = (\alpha_0^{(1)}, \alpha_1^{(1)}, \alpha_2^{(1)}, \lambda^{(1)})$, $\boldsymbol \theta^{(2)} = (\alpha_0^{(2)}, \alpha_1^{(2)}, \alpha_2^{(2)}, \lambda^{(2)})$, ...,$\boldsymbol \theta^{(n)} = (\alpha_0^{(n)}, \alpha_1^{(n)}, \alpha_2^{(n)}, \lambda^{(n)})$
\end{enumerate}

Finally, the Bayes estimates of the parameters with respect to squared error loss are given by
$$
\widehat{\alpha_0}_{(B)}=\frac{1}{n} \sum_{k=1}^{n} \alpha_0^{(k)}, \quad \widehat{\alpha}_{1(B)}=\frac{1}{n} \sum_{k=1}^{n} \alpha_1^{(k)}, \quad \widehat{\alpha}_{2(B)}=\frac{1}{n} \sum_{k=1}^{n} \alpha_2^{(k)}, \quad \widehat{\lambda}_{(B)}=\frac{1}{n} \sum_{k=1}^{n} \lambda^{(k)}.
$$
The posterior variances can be computed as
$$
V_{post}(\alpha_0)=\frac{1}{n} \sum_{k=1}^{n} (\alpha_0^{(k)} - \widehat{\alpha_0}_{(B)})^2, \quad V_{post}(\alpha_1)=\frac{1}{n} \sum_{k=1}^{n} (\alpha_1^{(k)} - \widehat{\alpha_1}_{(B)})^2 $$

$$V_{post}(\alpha_2)=\frac{1}{n} \sum_{k=1}^{n} (\alpha_2^{(k)} - \widehat{\alpha_2}_{(B)})^2, \quad V_{post}(\lambda)=\frac{1}{n} \sum_{k=1}^{n} (\lambda^{(k)} - \widehat{\lambda}_{(B)})^2.
$$

For obtaining Bayesian credible intervals for the parameter $\alpha_0$, first the posterior sample is ordered as follows: 
$$ \alpha_0^{(1)} < \alpha_0^{(2)} < ... < \alpha_0^{(n)}.$$
A $100(1-\gamma)\%$ Bayesian credible interval for $\alpha_0$ is then $(\alpha_0^{[\frac{\gamma}{2}{n}]}, \alpha_0^{[(1-\frac{\gamma}{2})n]})$, where $[x]$ is the symbol for greatest integer not exceeding $x$. Bayesian credible intervals for the other parameters can be obtained similarly. 

\section{\sc Monte Carlo simulation study}\label{sec:MC}
The motivation for the Monte Carlo simulation study is to examine performances of the MLEs and Bayes estimates of parameters of the MOBWDS model when applied to censored reliability data with two dependent failure modes. For the simulations, three different sample sizes are used: $n$ = 100, 200, and 400. Two different sets of values for the model parameters $\alpha_0$, $\alpha_1$, $\alpha_2$, and $\lambda$ are used. We evaluate the performance of the estimates through relative mean squared error (MSE) and relative bias, as defined below. If the true value of $\alpha_0$ is $\alpha_0^*$, then relative bias and relative MSE for the MLE $\widehat{\alpha}_0$ of $\alpha_0$ are calculated as follows: 
\begin{align*}
Relative \; MSE(\alpha_0) & = \frac{MSE (\widehat{\alpha}_0)}{(\alpha_0^*)^2} 
                          && \text{and} &
Relative \; Bias(\alpha_0) & = \frac{Bias(\widehat{\alpha}_0)}{\alpha_0^*}.
\end{align*}
The asymptotic CIs are assessed by their coverage probability and average length. The results of the simulation study are given in Tables \ref{table:wei-comp} - \ref{table:wei-cen2}. Relative bias and relative MSE of the Bayes estimates, as well as the coverage probability and average length of the credible intervals are also given in these tables.    

For the numerical optimization required to compute MLEs, the Nelder-Mead method, which is a direct search approach, is implemented through the \texttt{optim()} function in \texttt{R} software. For Bayes estimates, hyperparameters are taken to be $a = b = c_1 = c_2 = 0.005$ and $a_0 = a_1 = a_2 = 1.2$.  

\begin{table}[htb]
    \scriptsize
	\caption{\scriptsize Relative bias and MSE of the MLEs and Bayes estimates, and coverage probability (CP) and average length (AL) for the asymptotic confidence and credible intervals for the model parameters, estimated based on competing risks reliability data with no censoring when true model is MOBWDS($\alpha_0 = 1.63$, $\alpha_1 = 1.11$, $\alpha_2 = 1.92$, $\lambda = 2.35$). 
}
	\begin{center}
		\begin{tabular}{| c | c | c | c | c | c | c | c | c | c |}
			\toprule
			&  & \multicolumn{2}{c|}{\makecell{Point Estimate \\ (Frequentist)}} &    	
         	\multicolumn{2}{c|}{\makecell{95\% CI \\ (Frequentist)}} &
         	\multicolumn{2}{c|}{\makecell{Point Estimate \\ (Bayesian)}} &
         	\multicolumn{2}{c|}{\makecell{95\% CI \\ (Bayesian)}} \\
			\cmidrule(lr){3-4}\cmidrule(lr){5-6}\cmidrule(lr){7-8}\cmidrule(lr){9-10}
			Sample & Par. & Relative & Relative & AL & CP & Relative & Relative & AL & CP \\
			Size & & MSE & Bias &  &  & MSE & Bias &  &  \\ 
			\midrule
			\multirow{4}{*}{100}
			& $\alpha_0$ & 0.022 & 0.064 & 0.908 & 0.952 & 0.034 & 0.109 & 0.971 & 0.951 \\
			& $\alpha_1$ & 0.014 & 0.054 & 0.438 & 0.917 & 0.017 & 0.065 & 0.447 & 0.920 \\
			& $\alpha_2$ & 0.014 & 0.039 & 0.851 & 0.941 & 0.020 & 0.060 & 0.879 & 0.920 \\
			& $\lambda$ & 0.030 & 0.048 & 1.488 & 0.950 & 0.041 & 0.069 & 1.542 & 0.937 \\
			\midrule
			\multirow{4}{*}{200}
			& $\alpha_0$ & 0.012 & 0.059 & 0.642 & 0.939 & 0.017 & 0.076 & 0.660 & 0.912 \\
			& $\alpha_1$ & 0.007 & 0.044 & 0.308 & 0.920 & 0.009 & 0.047 & 0.311 & 0.898 \\
			& $\alpha_2$ & 0.007 & 0.030 & 0.597 & 0.940 & 0.009 & 0.032 & 0.603 & 0.926 \\
			& $\lambda$ & 0.013 & 0.029 & 1.030 & 0.956 & 0.016 & 0.036 & 1.048 & 0.950 \\
			\midrule
			\multirow{4}{*}{400}
			& $\alpha_0$ & 0.006 & 0.048 & 0.418 & 0.925 & 0.008 & 0.048 & 0.454 & 0.897 \\
			& $\alpha_1$ & 0.004 & 0.037 & 0.216 & 0.870 & 0.005 & 0.033 & 0.218 & 0.888 \\
			& $\alpha_2$ & 0.003 & 0.020 & 0.418 & 0.940 & 0.004 & 0.019 & 0.422 & 0.939 \\
			& $\lambda$ & 0.006 & 0.013 & 0.712 & 0.941 & 0.007 & 0.011 & 0.718 & 0.952 \\
			\bottomrule
		\end{tabular}
		\label{table:wei-comp}
	\end{center}
\end{table}

\begin{table}[htb]
    \scriptsize
	\caption{\scriptsize Relative bias and MSE of the MLEs and Bayes estimates, and coverage probability (CP) and average length (AL) for the asymptotic confidence and credible intervals for the model parameters, estimated from competing risks reliability data with approximately 20\% censoring when true model is MOBWDS($\alpha_0 = 1.63$, $\alpha_1 = 1.11$, $\alpha_2 = 1.92$, $\lambda = 2.35$).}
	\begin{center}
		\begin{tabular}{ | c | c | c | c | c | c | c | c | c | c |}
			\toprule
			&  & \multicolumn{2}{c|}{\makecell{Point Estimate \\ (Frequentist)}} &    	
         	\multicolumn{2}{c|}{\makecell{95\% CI \\ (Frequentist)}} &
         	\multicolumn{2}{c|}{\makecell{Point Estimate \\ (Bayesian)}} &
         	\multicolumn{2}{c|}{\makecell{95\% CI \\ (Bayesian)}} \\
			\cmidrule(lr){3-4}\cmidrule(lr){5-6}\cmidrule(lr){7-8}\cmidrule(lr){9-10}
			Sample & Par. & Relative & Relative & AL & CP & Relative & Relative & AL & CP \\
			Size & & MSE & Bias &  &  & MSE & Bias &  &  \\ 
			\midrule
			\multirow{4}{*}{100} 
			& $\alpha_0$ & 0.060 & 0.185 & 1.050 & 0.869 & 0.090 & 0.237 & 1.161 & 0.841 \\
			& $\alpha_1$ & 0.016 & 0.028 & 0.489 & 0.897 & 0.019 & 0.040 & 0.512 & 0.922 \\
			& $\alpha_2$ & 0.020 & 0.030 & 0.923 & 0.908 & 0.022 & 0.040 & 0.958 & 0.933 \\
			& $\lambda$ & 0.054 & 0.063 & 1.931 & 0.928 & 0.074 & 0.085 & 2.049 & 0.948 \\
			\midrule
			\multirow{4}{*}{200} 
			& $\alpha_0$ & 0.038 & 0.164 & 0.727 & 0.771 & 0.048 & 0.181 & 0.758 & 0.716 \\
			& $\alpha_1$ & 0.007 & 0.023 & 0.347 & 0.923 & 0.009 & 0.024 & 0.356 & 0.921 \\
			& $\alpha_2$ & 0.008 & 0.021 & 0.650 & 0.929 & 0.010 & 0.020 & 0.663 & 0.932 \\
			& $\lambda$ & 0.023 & 0.044 & 1.345 & 0.941 & 0.026 & 0.041 & 1.368 & 0.716 \\
			\midrule
			\multirow{4}{*}{400} 
			& $\alpha_0$ & 0.030 & 0.157 & 0.513 & 0.532 & 0.034 & 0.160 & 0.521 & 0.509 \\
			& $\alpha_1$ & 0.003 & 0.016 & 0.245 & 0.927 & 0.005 & 0.020 & 0.250 & 0.901 \\
			& $\alpha_2$ & 0.004 & 0.013 & 0.456 & 0.930 & 0.005 & 0.016 & 0.466 & 0.929 \\
			& $\lambda$ & 0.012 & 0.031 & 0.937 & 0.928 & 0.014 & 0.035 & 0.961 & 0.946 \\
			\bottomrule
		\end{tabular}
		\label{table:wei-cen1}
	\end{center}
\end{table}

\begin{table}[htb]
    \scriptsize
	\caption{\scriptsize Relative bias and MSE of the MLEs and Bayes estimates, and coverage probability (CP) and average length (AL) for the asymptotic confidence and credible intervals for the model parameters, estimated from competing risks reliability data with approximately 40\% censoring when true model is MOBWDS($\alpha_0 = 1.63$, $\alpha_1 = 1.11$, $\alpha_2 = 1.92$, $\lambda = 2.35$).}
	\begin{center}
		\begin{tabular}{ | c | c | c | c | c | c | c | c | c | c |}
			\toprule
			&  & \multicolumn{2}{c|}{\makecell{Point Estimate \\ (Frequentist)}} &    	
         	\multicolumn{2}{c|}{\makecell{95\% CI \\ (Frequentist)}} &
         	\multicolumn{2}{c|}{\makecell{Point Estimate \\ (Bayesian)}} &
         	\multicolumn{2}{c|}{\makecell{95\% CI \\ (Bayesian)}} \\
			\cmidrule(lr){3-4}\cmidrule(lr){5-6}\cmidrule(lr){7-8}\cmidrule(lr){9-10}
			Sample & Par. & Relative & Relative & AL & CP & Relative & Relative & AL & CP \\
			Size & & MSE & Bias &  &  & MSE & Bias &  &  \\ 
			\midrule
			\multirow{4}{*}{100} 
			& $\alpha_0$ & 0.083 & 0.246 & 1.126 & 0.817 & 0.123 & 0.297 & 1.279 & 0.770\\
			& $\alpha_1$ & 0.019 & 0.027 & 0.556 & 0.916 & 0.021 & 0.029 & 0.587 & 0.953\\
			& $\alpha_2$ & 0.025 & 0.038 & 1.006 & 0.908 & 0.026 & 0.036 & 1.047 & 0.923\\
			& $\lambda$ & 0.102 & 0.104 & 2.492 & 0.940 & 0.099 & 0.109 & 2.655 & 0.968\\
			\midrule
			\multirow{4}{*}{200} 
			& $\alpha_0$ & 0.062 & 0.225 & 0.773 & 0.581 & 0.076& 0.246& 0.818& 0.544 \\
			& $\alpha_1$ & 0.008 & 0.009 & 0.386 & 0.917 & 0.010& 0.011& 0.399& 0.939\\
			& $\alpha_2$ & 0.009 & 0.019 & 0.693 & 0.915 & 0.012& 0.022& 0.720& 0.918\\
			& $\lambda$ & 0.036 & 0.054 & 1.657 & 0.935 & 0.048& 0.067& 1.736& 0.954\\
			\midrule
			\multirow{4}{*}{400}
			& $\alpha_0$ & 0.052 & 0.215 & 0.541 & 0.239 & 0.057 & 0.218 & 0.599 & 0.266 \\
			& $\alpha_1$ & 0.004 & 0.002 & 0.271 & 0.920 & 0.005 & 0.013 & 0.280 & 0.930 \\
			& $\alpha_2$ & 0.004 & 0.009 & 0.487 & 0.929 & 0.006 & 0.001 & 0.504 & 0.928 \\
			& $\lambda$ & 0.017 & 0.038 & 1.148 & 0.933 & 0.022 & 0.040 & 1.189 & 0.933 \\
			\bottomrule
		\end{tabular}
		\label{table:wei-cen2}
	\end{center}
\end{table}

From Table \ref{table:wei-comp}, it may be observed that when there is no censoring in the data, the point and interval estimates of $\alpha_0$, $\alpha_1$, $\alpha_2$, and $\lambda$ are reasonably well. Relative bias and relative MSE reduce with increasing sample size, as one would expect. For interval estimates, although their average length reduces with increasing sample size as expected, the coverage probability still remains close to the nominal level of 95\%.

For censored data, as can be seen from Tables \ref{table:wei-cen1} and \ref{table:wei-cen2}, relative bias and relative MSE are once again reasonable, showing expected trends of reduction with increasing sample size. However, the intervals for $\alpha_0$ have poor coverage probability, although for the other parameters the coverage probability is close to the nominal confidence level. The problem of poor coverage of the intervals for $\alpha_0$ increases with increase in censoring percentage. The main reason for this seems to be the biased estimates of $\alpha_0$; even with increasing sample size, the bias of the estimate for $\alpha_0$ does not reduce much, and as a result, the narrowing intervals cannot retain the coverage probability. One possible solution for this could be the use of a standard bias reduction technique, such as the jackknife approach, to reduce bias in the estimates. Similarly, a bias-corrected bootstrap approach for calculating the intervals may also be used. 

\section{\sc Prediction of Future Failures} \label{sec:Pred}
Here we develop approaches for prediction of future failures based on censored reliability data with two dependent failure modes. First, prediction of future failures regardless of the failure mode is discussed. Then, prediction of future failures corresponding to a specific failure mode is considered. 

Suppose there are $n$ units in a reliability field or laboratory study. The lifetimes of the units under study are right censored at $R$. Let there be $n^*$ units that are right censored. Consider a future time interval that starts immediately after the end of the study; let $(R, R+\delta)$ where $\delta > 0$ denote such an interval. Our goal is to predict the number of failures out of the $n^*$ right censored units in this future interval $(R, R+\delta)$. 

\subsection{\sc Prediction of future failures regardless of failure mode} 
The lifetimes corresponding to the two dependent failure modes are denoted by $X$ and $Y$. The probability that a unit will fail in the future interval $(R, R+\delta)$ from any one of the two failure modes is given by
\begin{eqnarray}
& \rho & = P(\textrm{Min}\{X,Y\} < R+\delta \mid \textrm{Min}\{X,Y\} > R) \nonumber \\
&& = \frac{S_T(R) - S_T(R+\delta)}{S_T(R)} \nonumber \\
&& = 1 - \frac{S_T(R+\delta)}{S_T(R)}, \label{eq:pred-all}
\end{eqnarray}
where $T$ is the minimum of $X$ and $Y$, and $S_T(x) = P(T > x)$ is the survival function of $T$. For the MOBWDS distribution, we have 
\[
S_T(t) = P(\textrm{Min}\{X,Y\} > t) = e^{-\lambda(t^{\alpha_0}+t^{\alpha_1}+t^{\alpha_2})}, 
\]
and hence $\rho = \rho(\alpha_0, \alpha_1, \alpha_2, \lambda)$ depends on the MOBWDS model parameters.

Suppose $M$ be the random number of failures out of the $n^*$ right censored units in the future interval $(R, R+\delta)$. Note that $M$ has a binomial distribution,
\[
M \sim \textrm{Binomial}(n^*, \rho),
\]
with $n^*$ as the number of trials, and $\rho$ as the probability of `success' (which is failure of a right censored unit in the future interval $(R, R+\delta)$).  

We first estimate the probability of a failure in the interval $(R, R+\delta)$ by plugging-in the MLEs of the model parameters: $\widehat{\rho} = \rho(\widehat{\alpha}_0, \widehat{\alpha}_1, \widehat{\alpha}_2, \widehat{\lambda})$. Then, as a frequentist prediction of number of future failures, we obtain the expected number of failures in the future interval $(R, R+\delta)$ as 
\[
E(M) = n^*\widehat{\rho}.
\]
The lower and upper prediction bounds for the number of failures, for a prediction interval with $100(1-\gamma)$\% confidence, can be obtained to be $M_{lower}$ and $M_{upper}$, where $M_{lower}$ is the $\gamma_1$-th quantile and $M_{upper}$ is the $\gamma_2$-th quantile of Binomial$(n^*, \widehat{\rho})$ distribution, with $\gamma_1 + \gamma_2 = \gamma$.

Bayesian prediction is based on generations from the joint posterior distribution \cite{Lewis-Beck}. The predictive distribution of the random variable $M$ is given by
\begin{equation}
\pi_{predictive}(m) = \int_{\boldsymbol \theta}F_M(m;\rho)\tilde{\pi}\left(\boldsymbol \theta \mid \text {Data}\right) d\boldsymbol \theta, \label{eq:pred-dens}
\end{equation}
where $F_M(m;\rho) = \sum_{j=0}^{m}\binom{n^*}{j}\rho^j(1-\rho)^{n^*-j}$ is the CDF of the random variable $M$, and $\tilde{\pi}\left(\boldsymbol \theta \mid \text {Data}\right) = \tilde{\pi}\left(\alpha_0, \alpha_1, \alpha_2, \lambda \mid \text {Data}\right)$ is the joint posterior distribution. 

For calculation purposes, the predictive distribution $\pi_{predictive}(m)$ needs to be approximated based on generations from the joint posterior distribution. Suppose, $\boldsymbol \theta_1^*, \boldsymbol \theta_2^*, ...,\boldsymbol \theta_R^*$ are $R$ generations from the joint posterior distribution $\tilde{\pi}\left(\boldsymbol \theta \mid \text {Data}\right)$. Then the predictive distribution $\pi_{predictive}(m)$ is approximated as 
\[
\pi_{predictive}(m) = \frac{1}{R}\sum_{k=1}^R F_M(m;\rho_k^*)\tilde{\pi}\left(\boldsymbol \theta_k^* \mid \text {Data}\right),
\]
where $\rho_k^* = \rho(\boldsymbol \theta_k^*)$. A Bayesian prediction for the number of failures in the future interval $(R, R+\delta)$ can then be given by the mean or median of the approximated predictive distribution. A Bayesian $100(1-\gamma)$\% prediction interval for the number of failures in the interval $(R, R+\delta)$ is given by the $\gamma_1$- and $\gamma_2$-percentiles of the predictive distribution $\pi_{predictive}(m)$, with $\gamma_1 + \gamma_2 = \gamma$. Lewis-Beck et al.~\cite{Lewis-Beck} gave more approximation ideas of the predictive distribution based on simulations.  

\subsection{\sc Prediction of future failures for a specific failure mode}
For predicting number of failures from a specific mode in the future interval $(R, R+\delta)$, the probability of failure needs to be modified. The probability of a failure from Mode 1, for example, will be
\begin{eqnarray}
& \rho_1 & = P(X < Y, X \leq R+\delta \mid \textrm{Min}(X,Y) > R) \nonumber \\
&& = \frac{P(X < Y, X \leq R+\delta, X > R, Y > R)}{P(\textrm{Min}(X, Y)> R)} \nonumber \\
&& = \frac{1}{S_T(R)}\int_R^{R+\delta}\int_x^{\infty}f_{X,Y}(x,y)dydx. \label{eq:pred-mode1}
\end{eqnarray} 
Then, the random number of failures from Mode 1 in the future interval $(R, R+\delta)$, denoted by $M_1$, say, has a binomial distribution
\[
M_1 \sim \textrm{Binomial}(n^*, \rho_1). 
\]
The frequentist and Bayesian prediction of number of failures from Mode 1 in the future interval $(R, R+\delta)$ can then be obtained similarly as above, using $\widehat{\rho}_1 = \rho_1(\widehat{\alpha}_0, \widehat{\alpha}_1, \widehat{\alpha}_2, \widehat{\lambda})$ in the distribution of $M_1$. The prediction of failures from Mode 2 can be obtained similarly, by adjusting the probability of a failure from Mode 2 in the future interval $(R, R+\delta)$.     

\section{\sc Analysis of device failure data}\label{sec:Data-analysis}
The device failure data presented in Meeker and Escobar~\cite{Meeker} are analysed here as an illustration. The data have information on running times (in terms of thousands of cycles) and mode of failure for 30 units of a device that is part of a larger system. For each failed unit, the mode of failure was determined. There are two possible modes: Mode S which is a failure due to ``accumulation of randomly occurring damage from power-line voltage spikes during electric storms'', and Mode W which is a failure due to normal product wear. Out of the 30 units under study, 8 were right censored as they were still in operating condition at 300 thousand cycles. Among the failed units, 7 failures were caused by Mode W, and 15 failures were caused by Mode S. 

Suppose $X$ and $Y$ be the lifetimes corresponding to Mode S and Mode W, respectively. Clearly, the observed lifetime of a failed unit will be $T = min (X, Y)$. As $X$ and $Y$ can infleunce each other, it would not be appropriate to model $X$ and $Y$ marginally, without considering a dependence mechanism. We model the bivariate random vector $(X, Y)$ by the MOBWDS model, and estimate the model parameters based on the observed data on device failures from Mode S and Mode W. Table \ref{table:example} presents the model fitting results to the device failure data.  
\begin{table}[h]
	\scriptsize
	\caption{\scriptsize MLEs and 95\% confidence intervals, and Bayes estimates and 95\% credible intervals for the parameters of the MOBWDS model applied to device failure data with two failure modes.}
	\begin{center}
		\begin{tabular}{ |c |c c |c c|} 
			\toprule
			 & \multicolumn{2}{c|}{Likelihood Inference} & \multicolumn{2}{c|}{Bayesian Inference} \\
			 Parameter & Estimate & 95\% Confidence Interval & Estimate & 95\% Credible Interval\\
			 \midrule 
			
			$\alpha_0$ & 0.234 & (0, 1.072) & 0.528 & (0.079, 1.242)\\
			$\alpha_1$ & 2.070 & (1.104, 3.036) & 2.171 & (1.480, 2.970) \\
			$\alpha_2$ & 0.761 & (0.394, 1.127) & 0.818 & (0.519, 1.145) \\
			$\lambda$  & 0.180 & (0.090, 0.270) & 0.169 & (0.104, 0.251) \\
			\bottomrule
		\end{tabular}
		\label{table:example}
	\end{center}
\end{table}

\begin{figure}
\subfloat{\includegraphics[width=1\linewidth, height=.6\linewidth]{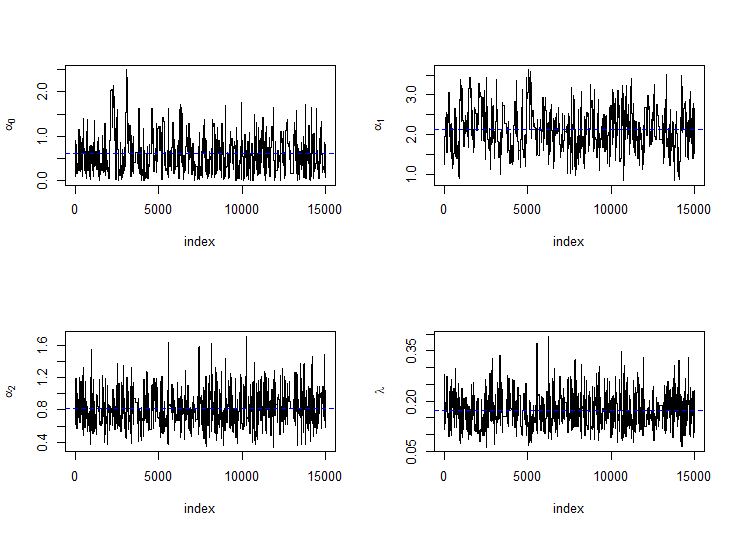}}
\caption{MCMC diagnostic: trace plot for $N$ = 15000.}
\label{mcmc}
\end{figure}

\begin{figure}
\subfloat{\includegraphics[width=1\linewidth, height=.6\linewidth]{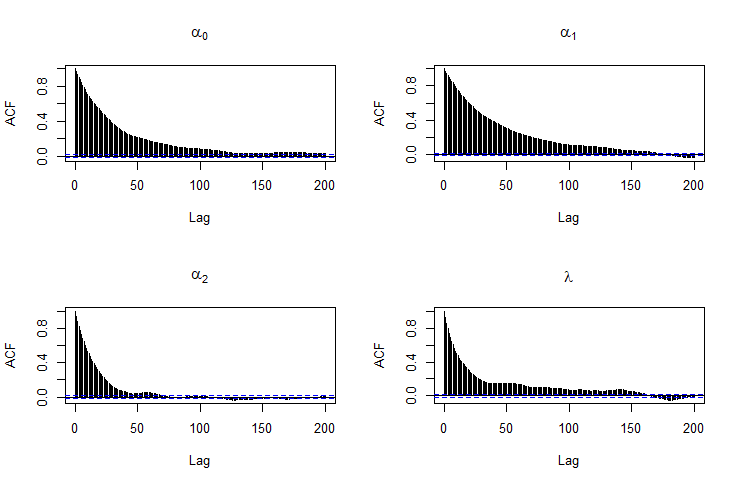}}
\caption{MCMC Diagnostic: autocorrelation plots}
\label{autocorr}
\end{figure}

For the Bayesian analysis of the device failure data, the hyper parameters are chosen to be $a = b = c_1 = c_2 = 0.005$ and $a_0 = a_1 = a_2 = 1.2$, thus using a diffuse prior. For implementing the Metropolis Hastings algorithm, we start with an arbitrary initial guess $\boldsymbol \theta = (1.40,0.95,1.85,0.60)$; chain of length $N = 15500$ is used, of which first 500 generations are discarded as burn-in. From the trace plot and autocorrelation plot given in Figures \ref{mcmc} and \ref{autocorr}, respectively, no apparent issue with the convergence to the target distribution is observed.  

Gelman and Rubin's~\cite{Gelman} MCMC convergence diagnostic measure that takes into consideration the between-chain and within-chain variances to assess convergence is also calculated. 
Based on the device failure data, we observe that the Gelman and Rubin's diagnostic statistic corresponding the four parameters are as follows: $\hat{R}_{\alpha_0} = 1.00332$; $\hat{R}_{\alpha_1} = 1.001001$; $\hat{R}_{\alpha_2} = 1.001507$; $\hat{R}_{\lambda} = 1.001622$. A numerical value of $\hat{R}_i$ close to one implies no issues with convergence. 

\begin{figure}
\centering
\subfloat{\includegraphics[width=0.5\linewidth, height=0.5\linewidth]{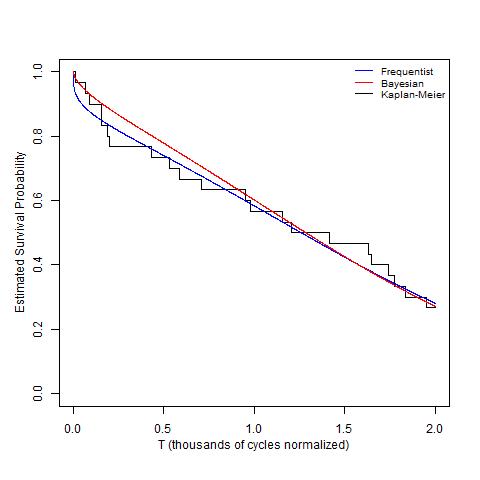}}
\caption{Comparison of the fitted Weibull survival function (the smooth curve) for bayesian and frequentist setup with the Kaplan-Meier curve (the step curve) for the Device Failure Data}
\label{kaplan-meier}
\end{figure}

In Figure \ref{kaplan-meier}, the nonparametric estimate of the survival function, known as the Kaplan-Meier estimate, is plotted along with the plots of the estimated survival curve assuming the MOBWDS model for the device failure data. For plotting the Kaplan-Meier estimate, we have considered only the failure time, ignoring the failure mode. Similarly, for plotting the parametric survival curves, the survival function of the minimum of the two failure times is used. It may be observed in Figure \ref{kaplan-meier} that the nonparametric and parametric estimates of the survival curve are quite close.         

The estimate of the mean time to failure (MTTF) is given by
\begin{equation}
\widehat{MTTF} = \int_0^{\infty}\widehat{S}(t,t)dt. \label{eq:mttf}
\end{equation}
For the device failure data, an estimated MTTF value under the MOBWDS model turns out to be $\widehat{MTTF}_{frequentist} = 210.27$ thousand of cycles, by plugging-in the MLEs in Eq.\eqref{eq:mttf} and then carrying out the subsequent calculations. For the Bayesian estimation of MTTF, the following steps are followed: 
\begin{enumerate}
[label=\arabic*.,itemsep=-1ex]
\item Generate $n$ samples $\boldsymbol \theta_1^*, \boldsymbol \theta_2^*,..., \boldsymbol \theta_n^*$ from the joint posterior distribution
\item For the $i$-th sample, calculate $\widehat{MTTF}_i$ using $\boldsymbol \theta_i^*$ in Eq.\eqref{eq:mttf}, $i=1,...,n$\\
\item The Bayes estimate of MTTF is finally calculated as 
\[
\widehat{MTTF}_{Bayes} = \frac{1}{n}\sum_{i=1}^{n}\widehat{MTTF}_i.
\]
\end{enumerate}
Based on 15000 samples from the joint posterior distribution, the Bayes estimate of MTTF is obtained for the device failure data; it turns out to be $\widehat{MTTF}_{Bayes} = 219.68$. 

\subsection{\sc Prediction of future failures: Device failure data}
In the device failure data, there are 8 right censored units, all of which are right censored at 300 thousand cycles. In Table \ref{table:prediction_ign}, the predicted number of failures regardless of the failure mode in future intervals of different lengths are given, along with corresponding prediction bounds for the predicted number of failures. Predicted number of failures with corresponding prediction bounds specific to a failure mode are given in Tables \ref{table:predictionx} and \ref{table:predictiony} for Mode S and Mode W failure, respectively. 

It may be observed that the predictions by frequentist and Bayesian methods are in close agreement. Also, as can be seen in the data, there are more Mode S failures compared to the Mode W failures. In mode specific prediction of failures, this trend is reflected: predicted number of failures and their corresponding prediction bounds for Mode S are significantly larger than their counterparts for Mode W.      
\begin{table}[h]
	\scriptsize
	\caption{\scriptsize Prediction of future failures regardless of failure mode for the device failure data, choosing $\delta = 30, 75, 120, 200$.}
	\begin{center}
		\begin{tabular}{ |c |c c |c c|} 
			\toprule
			 & \multicolumn{2}{c|}{Frequentist Prediction} & \multicolumn{2}{c|}{Bayesian Prediction} \\
			 Future Interval & Predicted & 95\% Prediction Bound & Predicted & 95\% Prediction Bound \\
			 & Number of Failures &  & Number of Failures & \\
			 \midrule 
			(300, 330) & 1.40 & (0, 3) & 1.59 & (0, 4)\\
			(300, 375) & 3.20 & (1, 5) & 3.51 & (0, 6) \\
			(300, 420) & 4.63 & (2, 7) & 4.91 & (1, 8) \\
			(300, 500)  & 6.37 & (4, 8) & 6.42 & (3, 8) \\
			\bottomrule
		\end{tabular}
		\label{table:prediction_ign}
	\end{center}
\end{table}

\begin{table}[h]
    \scriptsize
	\caption{\scriptsize Prediction of future failures from Mode S for the device failure data, choosing $\delta = 30, 75, 120, 200$.}
	\begin{center}
		\begin{tabular}{ |c |c c |c c|} 
			\toprule
			 & \multicolumn{2}{c|}{Frequentist Prediction} & \multicolumn{2}{c|}{Bayesian Prediction} \\
			 Future Interval & Predicted & 95\% Prediction Bound & Predicted & 95\% Prediction Bound \\
			 & Number of Failures &  & Number of Failures & \\
			 \midrule 
			(300, 330) & 1.18 & (0, 3) & 1.26 & (0, 3)\\
			(300, 375) & 2.72 & (1, 5) & 2.82 & (0, 6) \\
			(300, 420) & 3.99 & (2, 6) & 3.97 & (0, 7) \\
			(300, 500)  & 5.55 & (3, 8) & 5.25 & (1, 8) \\
			\bottomrule
		\end{tabular}
		\label{table:predictionx}
	\end{center}
\end{table}

\begin{table}[h]
	\scriptsize
	\caption{\scriptsize Prediction of future failures from Mode W for the device failure data, choosing $\delta = 30, 75, 120, 200$.}
	\begin{center}
		\begin{tabular}{ |c |c c |c c|} 
			\toprule
			 & \multicolumn{2}{c|}{Frequentist Prediction} & \multicolumn{2}{c|}{Bayesian Prediction} \\
			 Future Interval & Predicted & 95\% Prediction Bound & Predicted & 95\% Prediction Bound \\
			 & Number of Failures &  & Number of Failures &  \\
			 \midrule 
			(300, 330) & 0.03 & (0, 0) & 0.05 & (0, 1)\\
			(300, 375) & 0.08 & (0, 1) & 0.14 & (0, 1) \\
			(300, 420) & 0.13 & (0, 1) & 0.20 & (0, 1) \\
			(300, 500)  & 0.20 & (0, 1) & 0.28 & (0, 1) \\
			\bottomrule
		\end{tabular}
		\label{table:predictiony}
	\end{center}
\end{table}


\section{\sc Concluding remarks}\label{sec:Conclusion}

In this article, modelling of censored reliability data with two dependent failure modes is discussed by using a proposed bivariate Weibull distribution, called the Marshall-Olkin bivariate Weibull model with distinct shape parameters. Likelihood and Bayesian inference for this issue are studied in detail. Through a Monte Carlo simulation study, it is observed that the proposed model and methods perform reasonably well. A relevant and practical prediction issue that involves prediction of future failures in a future interval, regardless of the failure modes as well as specific to failure modes, is discussed. Analysis of a real data on device failure with two failure modes is provided for illustrative purposes.   

Naturally, it would be of interest to extend this research to a multivariable setting where more than two failure modes can compete for failures. An extension of the bivariate model considered here to the multivariable scenario may be considered in such a case. Marshall and Olkin~\cite{Marshall67} presented an outline for such an extension in case of a multivariate exponential distribution. Along with the usual limitations of the exponential distributions as marginal models for lifetimes, it has also been observed that in case of the multivariate exponential distribution, the analytical treatment gets quite complicated due to the existence of singular distributions in lower dimensions. An extension to the multivariable case for Weibull distribution with distinct shape parameters would be useful in reliability in modelling reliability data with multiple failure modes. But such an extension to a multivariate distribution is not straightforward. This would be an interesting problem for future research.

\section*{\sc Declaration of conflict of interest}
The authors declare that there is no conflict of interest.

\section*{\sc Funding information}
The research of Ayon Ganguly is supported by the Mathematical Research Impact
Centric Support (File no.~MTR/2017/000700) from the Science and
Engineering Research Board, Department of Science and Technology, Government of
India. \\
Debanjan Mitra thanks Indian Institute of Management Udaipur for financial
support to carry out this research.


\singlespacing


\begin{thebibliography}{99}


\bibitem{Abba}
Abba, B., Wang, H., and Bakouch, H. S. (2022). A reliability and survival model for one and two failure modes system with application to complete and censored datasets. {\it Reliability Engineering \& System Safety}, {\bf 223}, \url{https://doi.org/10.1016/j.ress.2022.108460}

 	
\bibitem{Almalki}
Almalki, S. J., and Yuan, J. (2013). A new modified Weibull distribution. {\it Reliability Engineering System \& System Safety}, {\bf 111}, 164--170. 

\bibitem{Cox}
Cox, D. R. (1959). The analysis of exponentially distributed lifetimes with two types of failures. {\it Journal of the Royal Statistical Society Series B}, {\bf 21}, 411--421.

\bibitem{Crowder} 
Crowder M. (2001). {\it Classical Competing Risks Model}. Chapman \& Hall: New York.	

\bibitem{Escobar}
Escobar, L. A., and Meeker, W. Q. (1999). Statistical prediction based on censored life
data. {\it Technometrics}, {\bf 41}, 113--124. 

\bibitem{Fan} 
Fan, T. H., and Hsu, T. M. (2015). Statistical inference of a two-component series system with correlated log-normal lifetime distribution under multiple type-I censoring. {\it IEEE Transactions on Reliability}, {\bf 64}, 376--385.

\bibitem{Fan-X}
Fan, X., Wang, X., Zhang, X., and Yu, X. (2022). Machine learning based water pipe failure prediction: The effects of engineering, geology, climate and socio-economic factors. {\it Reliability Engineering \& System Safety}, {\bf 219}, \url{https://doi.org/10.1016/j.ress.2021.108185}


\bibitem{Feiz2015}
Feizjavadian, S. H., and Hashemi, R. Analysis of dependent competing risks in the presence of progressive hybrid censoring using Marshall–Olkin bivariate Weibull distribution. {\it Computational Statistics and Data Analysis} {\bf 82}, 19--34.

\bibitem{Gelman}
Gelman, A., and Rubin, D. B. (1992). Inference from Iterative Simulation Using Multiple Sequences. {\it Statistical Science}, {\bf 7}, 457--511. 

\bibitem{He}
He, B., Cui, W., and Du, X. (2016). An additive modified Weibull distribution. {\it Reliability Engineering \& System Safety}, {\bf 145}, 28--37.

\bibitem{Kundu-EM}
Kundu, D., and Dey, A. K. (2009). Estimating the parameters of the Marshall–Olkin bivariate Weibull distribution by EM algorithm. {\it Computational Statistics \& Data Analysis}, {\bf 53}, 956--965.

\bibitem{Kundu-Bayes}
Kundu, D., and Gupta, A. K. (2013). Bayes estimation for the Marshall–Olkin bivariate Weibull distribution. {\it Computational Statistics \& Data Analysis}, {\bf 57}, 271--281.


\bibitem{Lawless} 
Lawless, J. F. (1982). {\it Statistical Models and Methods for Lifetimes Data}. Wiley, New York.

\bibitem{Lewis-Beck} 
Lewis-Beck, C., Tian, Q., and Meeker, W. Q. (2021). Prediction of future failures for heterogeneous reliability field data. {\it Technometrics}, {\bf 64}, 125--138.


\bibitem{Marshall67}
Marshall, A. W., and Olkin, I. (1967). A multivariate exponential distribution. {\it Journal of the American Statistical Association}, {\bf 62}, 30--44.

\bibitem{Meeker}
Meeker, W. Q., and Escobar, L. A. (1998). {\it Statistical Methods for Reliability Data}, John Wiley \& Sons: New York.

\bibitem{Oliviera}
Oliviera, R. P., Achcar, J. A., Mazucheli, J., and Bertoli, W. (2021). A new class of bivariate Lindley distributions based on stress and shock models and some of their reliability properties. {\it Reliability Engineering \& System Safety}, {\bf 211}, \url{https://doi.org/10.1016/j.ress.2021.107528}

\bibitem{Pena}
Pena, E. A., and Gupta, A. K. (1990). Bayes estimation for the Marshall-Olkin exponential
distribution, {\it Journal of the Royal Statistical Society , Ser B}, {\bf 52}, 379--389.


\bibitem{Rifaai} 
Rifaai, T. M., Abokifa, A. A., and Sela, L. (2022). Integrated approach for pipe failure prediction and condition scoring in water infrastructure systems. {\it Reliability Engineering \& System Safety}, {\bf 220}, \url{https://doi.org/10.1016/j.ress.2021.108271} 

\bibitem{Samanta}
Samanta, D., and Kundu, D. (2021). Bayesian inference of a dependent competing risk data. {\it Journal of Statistical Computation and Simulation}, {\bf 91}, 3069--3086.

\bibitem{Xie}
Xie, M., and Lai, C. D. (1996). Reliability analysis using an additive Weibull model with bathtub-shaped failure rate function. {\it Reliability Engineering \& System Safety}, {\bf 52}, 87--93.

\end{thebibliography}
\end{document}